\documentclass[preprint,aps]{revtex4-2}
\pdfoutput=1
\usepackage{dcolumn}  % mathcalign table columns on decimmathcal point
\usepackage{bm}  % bold math
\usepackage{graphicx}% Include figure files
\usepackage{amsmath}
\usepackage{xcolor}

\graphicspath{{figs/gauss/}}
\begin{document}

\title{The sub - diffusive behavior of chromatin in changing environment}
%\author{Yevgeni Sh. Mamasakhlisov$^{1}$, Vladimir V Papoyan$^{2}$}
%\pacs{87.10.Ca, 87.15.hp, 87.14.gk, 82.37.Rs}
	  \author{Vladimir V.\ Papoyan}
	\email{vpap@theor.jinr.ru}
	\affiliation{BLTP, JINR,  Dubna,  Russian Federation}
	\affiliation{Dubna State University, Dubna, Russian Federation}
	\author{Yevgeni Sh.\ Mamasakhlisov}
	\email{y.mamasakhlisov@gmail.com}
	\affiliation{Yerevan State University, Yerevan, Armenia }
	\affiliation{Russian - Armenian University, Yerevan, Armenia }
	
\begin{abstract}
The rotor-router walk model is proposed to describe the sub - diffusive behavior of double - strand breaks in chromatin 
caused by external factors such as heavy ions, $\gamma$ - irradiation, etc.
The mechanism of the double - strand breaks reparation is considered in terms of this model.
\end{abstract}

\maketitle

\section{introduction}

The physical features of double- and single - stranded nucleic acids (DNA, RNA), such as elasticity, rigidity, and conformational mobility are well known \cite{r1, r2} while the dynamic and thermodynamic properties of chromatin {\it in vivo} still remain unclear. Not only has it been difficult to visualize specific chromatin loci, but it remains unclear what
physical properties should be measured. Till now we do not have a satisfactory physical physical model to interpret experimental data concerning chromatin dynamics and structure. However, recent works have advanced chromatin models to the point that they can predict changes in organization
 and dynamics \cite{shukron_2019_cell}

Chromatin motion is usually considered in terms of single locus trajectories and the main physical observable is 
the mean square displacement (MSD) that is defined as the squared displacement with respect to the initial locus position, averaged over time for a given loci $MSD(t)=\langle(X(t)-X(0))^2\rangle$. From this, a volume of space explored by the locus can be extracted.

To interpret the results of this kind of experiments a various physical diffusion or polymeric models are used. The model chosen
to fit an MSD curve must provide plausible physical mechanisms and accurate predictions.
Examples of physical models include Brownian motion or random walk and anomalous diffusion. As is was shown experimentally, the locus motion
is described as sub diffusive, where the $MSD\sim t^{\alpha}$ and $\alpha < 1$ \cite{shukron_2019_cell}.
The exponent $\alpha$ describing the dynamics of chromatin loci varies
in the range $0.3-2$ in dependence on various external factors \cite{cabal, backlund, bronstein}. 

The similar sub - diffusive behavior is observed for the dynamics of double - stranded breaks arising upon irradiation of chromatin by heavy ions. In accordance with current models on biomolecular condensation and liquid-liquid phase separation \cite{banani_2017},
upon nucleation by upstream events triggered at sites of DNA damage (cH2AX formation, MDC1 recruitment, activation of RNF8 and RNF168), 53BP1 accumulates and phase separates, acting as scaffold for client molecules such as p53, which transiently interact with the 53BP1 compartment, where they find an environment permissive for their
activation. As clients, with lower relative enrichment and higher mobility, they can dissociate again upon activation to carry out effector functions at distant sites in the
nucleus. Phase separation of 53BP1 may also promote compartment fusions and clustering of break sites, with potential implications for mis-rejoining of DNA breaks and
generation of chromosomal translocation.

Thus, the double - strand breaks (DSBs) formation results to the local changes in the environment affecting dynamics of chromatin. The present paper is focused on the general model describing how the environment local reorganization results to the sub - diffusive behavior of the DSBs.

\section{The model of changing environment}

Our goal is to describe reparation of DSB in terms of rotor-router walk (RRW) model. We are focused on the two main properties of the DSBs. First of all, DSBs exhibit sub diffusion behavior  and second, induced the environment reorganization process powered by very complicated biochemical reactions.

The RRW model shared with DSB the sub diffusive walk and at the same time reorganization of the medium  around.

%The remarkable feature of DSBs is their ability to change the environment and sub diffusive behavior.

\begin{figure}[htbp]
\centerline{\includegraphics[height=110pt]{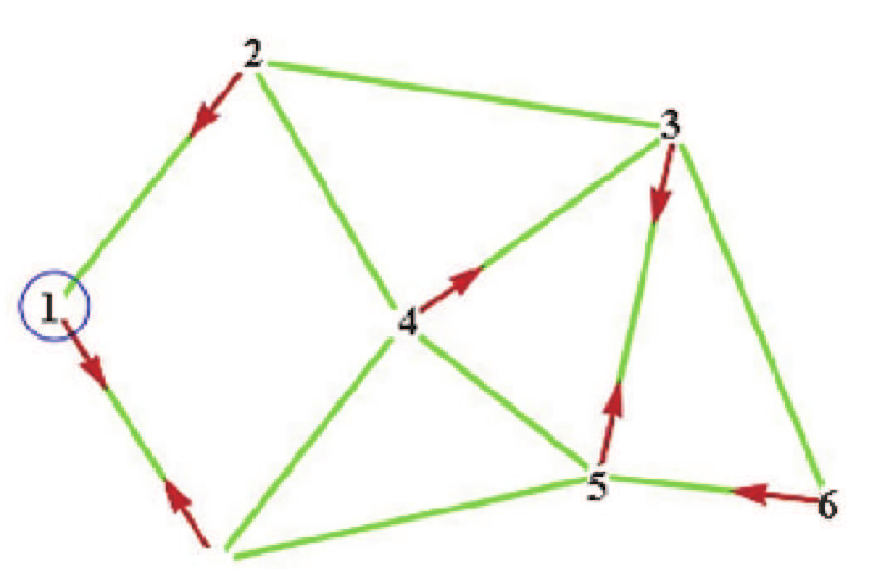}}%[width=3.63in,height=3.32in]
\caption{The rotor-router on directed graph. The configuration of arrows is initial one. The sequence of the visited sites is numbered.}
\label{fig_1}
\end{figure}

The model of the RRW can be defined as follows. Consider a directed graph with
arrows attached to vertices. The arrow in every vertex is directed to one of its neighbors
on the graph. A particle called usually chip performs a walk jumping from a vertex to a
neighboring vertex. Arriving to a given vertex, the chip changes direction of the arrow in this
site in a prescribed order and moves to the neighbor pointed by new position of the arrow Fig.\ref{fig_1}. The order of the arrow's direction change can be arbitrary but should be the same for any vertex of the graph during the whole time of motion of chip. Thus, given an initial orientation of arrows on the whole graph, the RRW is quite deterministic.

The rotor-router walk is the latter and most frequently used name of the model introduced independently by researchers in different communities during the last two decades. The previous names Self-directing walk first introduced by V.B. Priezzhev (1996) \cite{Pr} and  Eulerian walkers by V.B. Priezzhev, D. Dhar, et al (1996) \cite{PrDh}
reflected its connection with the theory of self-organized criticality [ P. Bak, C. Tang, and K. Wiesenfeld (1987) \cite{SOC} and the Abelian sandpile model \cite{Dh}. Cooper and Spencer (2006) \cite{CoSp} called the model P-machine after Propp (2001) \cite{Propp} who proposed the rotor mechanism as the way to derandomize models of random walks such as the internal diffusion-limited aggregation.

The rotor-router walk on a graph is a discrete-time walk according to the sub-diffusion law:
\begin{equation}
\langle r^2\rangle\sim t^{\frac{d}{d+1}},
\label{sub}
\end{equation}
where $r$ is the chip displacement, $t$ is the number of steps, and $d$ is the system dimensionality \cite{FlLePe}.
The motion of the chip is accompanied by a deterministic evolution of arrow configurations randomly placed on the vertices of the graph. The deterministic change of the graph properties describes the environment evolution.

We propose the simplified model, where the open end of DSB is identified as a chip of the RRW model and environment properties as arrow configuration of the RRW model.

\section{The dynamics of RRW}

\begin{figure}[htbp]
\centerline{\includegraphics[height=110pt]{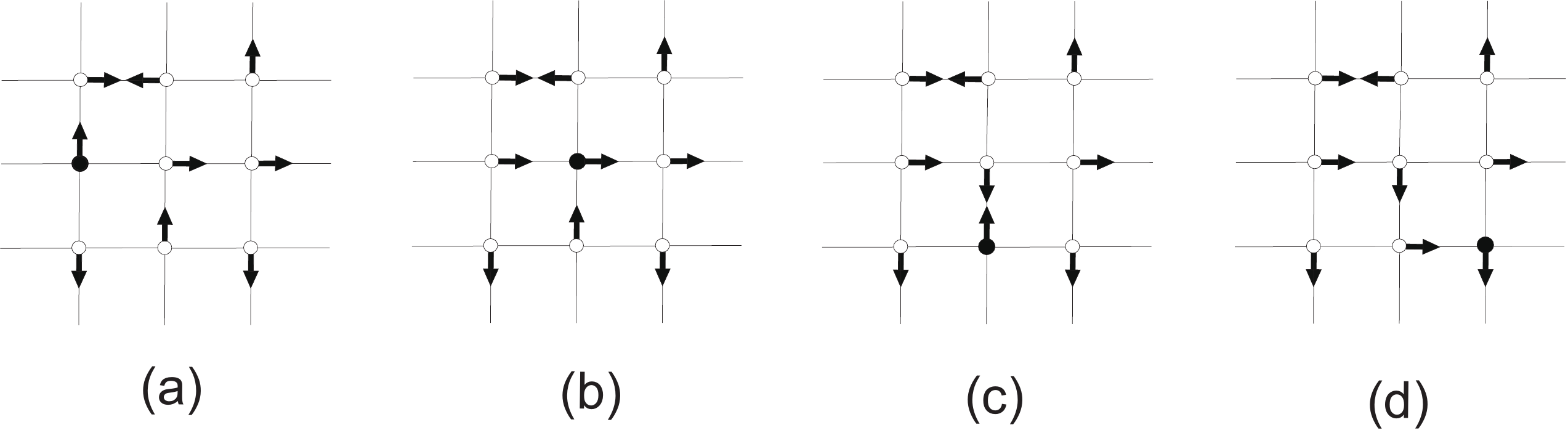}}%[width=3.63in,height=3.32in]
\caption{Circles denote the lattice sites. (a) The chip is originally in the filled
circle where the arrow is directed ‘up’. (b) The chip rotates the arrow clockwise
and moves right. (c) The next clockwise rotation sends the chip down. (d) The last
position of the chip is the lower right corner.}
\label{fig_2}
\end{figure}

For simplicity we will restrict ourselves by consideration of the RRW on the infinite square lattice with clockwise rotor mechanism at each site. In the initial rotor state, the arrows at each lattice site are directed randomly in one of four directions with equal probabilities, and the chip is at the origin. At each step of discrete time, the chip arriving at a site rotates the arrow at that site 90 degrees clockwise, and moves to the neighboring site pointed out by the new position of the arrows.
 Three steps of the RRW on the square lattice are shown in Fig.\ref{fig_2}.

\begin{figure}[htbp]
\centerline{\includegraphics[width=90mm]{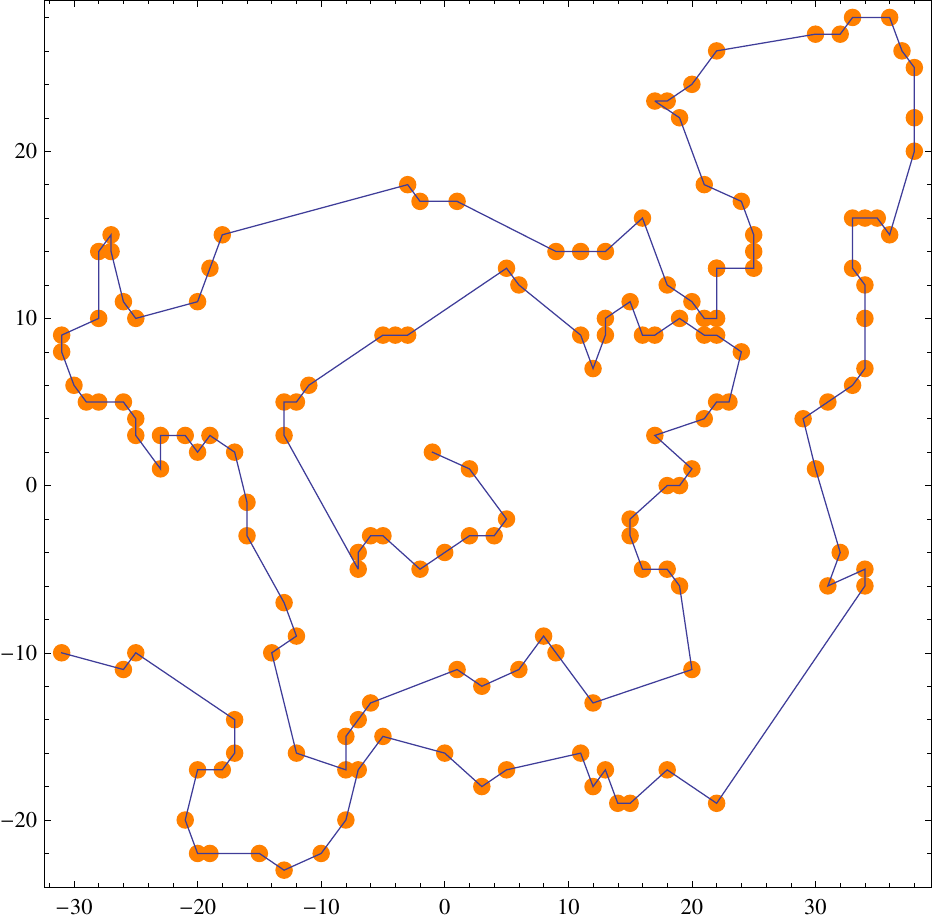}}%[width=3.63in,height=3.32in]
\caption{The spiral of nodes. The realization of $10^3$ steps of chip.}
\label{fig_3}
\end{figure}

To shed light onto the connection between the environment reorganization and sub-diffusion law we need to give some definitions of RRW following the authors of \cite{our16}.
In \cite{our15} the motion of the clockwise RRW inside closed contours emerged in random rotor configurations on the infinite square lattice has been considered. And the property called weak reversibility is proved: even though the configuration of rotors inside the contour is random, the RRW inside the contour demonstrates some regularity, namely, the chip entering the clockwise contour $\mathcal{C}$ in a vertex $v \in \mathcal{C}$ leaves the contour at the same vertex $v$, and then the clock-wise orientation of rotors on $\mathcal{C}$ becomes anti-clockwise. These vertices $v$ are called labels.
We will also need to consider specific labels called nodes.
Let $\mathcal{C}_i$ and $\mathcal{C}_{i+1}$ be two successive contours in the sequence $\mathcal{C}_1$, $\mathcal{C}_2$ ... There are
three possibilities for the disposition of $\mathcal{C}_{i+1}$ with respect to $\mathcal{C}_{i}$: (a) the set of sites $\{v\}_{\mathcal{C}_{i +1}}$ where arrows of $\mathcal{C}_{i+1}$ are attached has no sites in common with $\{v\}_{\mathcal{C}_i}$ and contour $\mathcal{C}_{i+1}$ is outside $\mathcal{C}_i$; (b) the set $\{v\}_{\mathcal{C}_{i +1}}$ has no common sites with $\{v\}_{\mathcal{C}_{i}}$ and contour $\mathcal{C}_{i}$ is inside $\mathcal{C}_{i+1}$; (c) the set $\{v\}_{\mathcal{C}_{i+1}}$ has at least one common site with $\{v\}_{\mathcal{C}_i}$ .{ To provide for condition (b), the contour $\{v\}_{\mathcal{C}_{i+1}}$  should contain sites visited at moments $t \leq t_i$ inside at the moment $t_{i+1}$. Otherwise, there are lattice sites outside $\{v\}_{\mathcal{C}_{i+1}}$  which do not connect with  at the moment $t_i$ by any path of arrows which is impossible for a single walk. When the cluster of visited sites grows, the probability of a contour enveloping the cluster of previously visited sites dramatically decreases and we can exclude case (b) from consideration.} Then, for a subset of labels $v_{i_1}$, $v_{i_2}$, ..., whose contours obey criterion (c) from the set of labels $v_{1}$, $v_{2}$, ...,  assume that $v_{i_1}$ coincides with $v_{1}$. The selected labels called "nodes". The sequence of nodes forms a spiral structure, see Fig.\ref{fig_3}. If the spiral is random, then it can be assumed that it asymptotically tends to the Archimedean spiral on average taken over all initial configurations of the rotors.

\begin{figure}[htbp]
\centerline{\includegraphics[width=80mm]{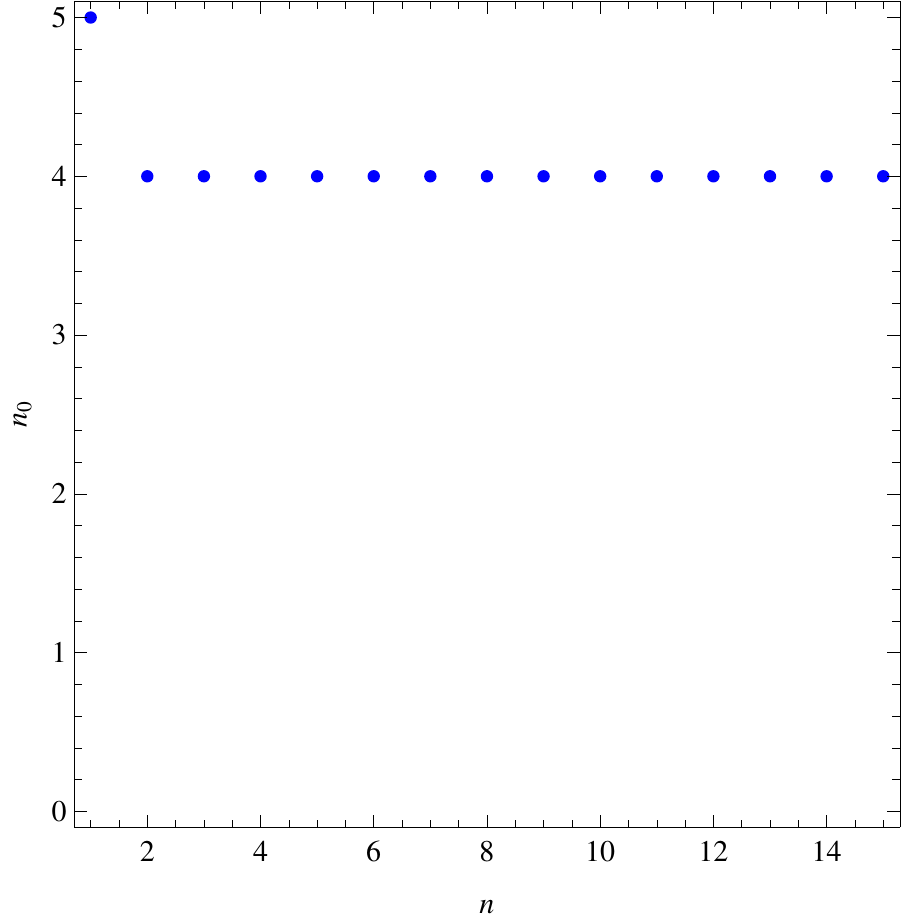}}%[width=3.63in,height=3.32in]
\caption{The number of visits to the origin $n_0$ at each turn of the spiral, depending on the number of turns n.}
\label{fig_4}
\end{figure}

We will also need the following important property RRW.
The number of visits to the origin of $N_0$ during the walk depends on the number of turns of the spiral $n$ as
\begin{equation}
N_0= 4 n+ O(1)
\label{number}
\end{equation}
where $O(1)$ is due to possible visits to the origin before the moment when the first loop
of the spiral is formed. In Fig.\ref{fig_4} it is shown that the number of visits to the origin $n_0$ at each turn of the spiral is exactly equal to $4$ after the formation of the spiral. The most surprising thing is that this result is true for any RRW even without averaging over the initial configurations.

\section{Results and discussion}

The effective reparation of the DSBs requires the close proximity of the broken ends of chromatin. The RRW model provides this feature through the mechanism of regular returns to the origin described above. Another important point of the reparation is the environment reorganized by chip. The DSB induced the self - organization of proteins and other molecules around the break according to the standard scenario involving the typical set of reparation proteins. The DNA damage triggered cH2AX formation, MDC1 recruitment, activation of RNF8 and RNF168 with the subsequent 53BP1 accumulation and phase separation, acting as scaffold for client molecules such as p53, which transiently interact with the 53BP1 compartments. We describe this phenomena in terms of regular reorganization of the environment in the RRW model.

In the RRW model the event signalling about the appearance of the ordered medium is the transition from the random walk on the initially random configuration of arrows to sub - diffusion.
At the same time the sub-diffusive behavior of the RRW model is inherently connected with the regular character of returns of the RRW chip as is was shown in \cite{our15}. In the context of DSBs reparation problem the return of the chip to the origin is necessary condition of reparation. At the same time, reparation process requires substantial reorganization of the environment, induced by DSBs. We propose the RRW model as a simplest way to describe the effect of this reorganization on the dynamics of DSB. Thus, the experimentally observed sub-diffusive behavior of the DSBs is the consequence of ordering of the environment around the DSBs.

The close proximity of the broken ends of chromatin is a necessary but not the only condition.
Besides of the close proximity we also need the appropriate state of the environment and the ends we need to repair. We propose to describe this state in terms of RRW model as a initial orientation of arrow in the origin of the walk, because the initial state of arrow corresponds to the unbroken state of the chromatin molecule. In order to get this original orientation is necessary to make four turns of the arrow. One turn of the spiral described in Fig.(\ref{fig_3}) results to  the four visits to the origin according to the Equation (\ref{number}) and Fig.(\ref{fig_4}). On the square lattice the four visits to the origin results to the four turns of the arrow.
Thus, in terms of RRW model we can identify the repaired ends of chromatin as a state with initial configuration of arrow at the origin of walk obtained as a result of return to the origin.

\end{document}